\documentclass[english]{article}
\usepackage[T1]{fontenc}
\usepackage[latin9]{inputenc}
\usepackage{amsmath}
\usepackage{amssymb}

\newcommand{\lyxaddress}[1]{
\par {\raggedright #1
\vspace{1.4em}
\noindent\par}
}

\usepackage{babel}

\begin{document}

\title{A COSMOLOGICAL MODEL OF THERMODYNAMIC OPEN UNIVERSE .}

\author{G.K.Goswami$^{1,3}$ \& Mandwi Trivedi$^{2}$}

\maketitle

\lyxaddress{\begin{center}
$^{\text{1}}$Kalyan PG College, 
\par\end{center}}

\lyxaddress{\begin{center}
Bhilai-490006 
\par\end{center}}

\lyxaddress{\begin{center}
(C.G.) INDIA
\par\end{center}}

\lyxaddress{\begin{center}
$^{\text{2}}Sri$Sankaracharya Engineering college
\par\end{center}}

\lyxaddress{\begin{center}
Gunvani Durg 
\par\end{center}}

\lyxaddress{\begin{center}
(C.G.) INDIA
\par\end{center}}

\lyxaddress{\begin{center}
$^{3}$E mail: gk\_goswami@yahoo.co.in
\par\end{center}}
\begin{abstract}
In this paper we have given a generalisation of the earlier work by
Prigogine et al$^{1-3}$. who have constructed a phenomenological
model of entropy production via particle creation in the very early
universe generated out of the vacuum rather than from a singularity,by
including radiation also as the energy source and tried to develop
an alternative cosmological model in which particle creation prevents
the big bang . We developed Radiation dominated model of the universe
which shows a general tendency that (i) it originates from instability
of vacuum rather than from a singularity. (ii) Upto a characteristic
time $t_{c}$ cosmological quantities like density, pressure, Hubble
constant and expansion parameter vary rapidly with time. (iii) After
the characteristic time these quantities settles down and the models
are turned into de-sitter type model with uniform matter, radiation,
creation densities and Hubble's constant H.The de-sitter regime survives
during a decay time $t_{d}$ then connects continuously to a usual
adiabatic matter radiation RW universe.\textbf{The intersesting thing
in the paper is that we have related the phenomenological radiation
dominated model to macroscopic model of quantum particle creation
in the early universe giving rise to the present observed value of
cosmic bachground radiation }. It is also found that the dust filled
model tallies exactly with that of the Prigogine's one, which justifies
that our model is generalized Prigogine's model.\textbf{Although the
model originates from instability of vacuum rather than from a singularity,still
there is a couple of unavoidable singularities in the model.} 
\end{abstract}

\section{{\large Introduction}\textmd{\large :}}

Prigogine and others$^{1-3\text{ }}$have proposed a phenomenological
model of the universe which originates from the instability of vacuum
rather than a singularity. For this they considered the whole universe
as an open thermodynamical system which modifies the usual adiabatic
energy conservation law leading to large scale entropy production
and possible irreversible matter creation in the universe. The model
which they presented was dust filled.There has been a lot of activity
in phenomenological models of cosmological particle production (see,
e.g.,$[4-11]$ to mention just a few). 

In this paper we have considered the similar system and have assumed
that entropy in the universe varies exponencially with time.We developed
Radiation dominated model of the universe which shows a general tendency
that 

(i) it originates from instability of vacuum rather than from a singularity. 

(ii) Upto a characteristic time $t_{c}$ cosmological quantities like
density, pressure, Hubble constant and expansion parameter vary rapidly
with time. 

(iii) After the characteristic time these quantities settles down
and the models are turned into de-sitter type model with uniform matter,
radiation, creation densities and Hubble's constant H.The de-sitter
regime survives during a decay time $t_{d}$ then connects continuously
to a usual adiabatic matter radiation RW universe.

\textbf{The intersesting thing in the paper is that we have related
the phenomenological radiation dominated model to macroscopic model
of quantum particle creation in the early universe giving rise to
the present observed value of cosmic bachground radiation . It is
also found that the Prigogine' model can be obtained from the field
equations which justifies that our model is generalized Prigogine's
model.}

A model, similar to that of the present manuscript, has been dealt
in ref$\left[4\right]$ . The main difference to the present study
is that Gunzig et al. used a variable particle production rate which
allowed them to describe the transition from an early de Sitter phase
to a subsequent radiation epoch analytically as a smooth process where
as the present paper works with a constant production rate. \textbf{As
regards the comparison of our results with those of ref. $\left[4\right]$,}
\textbf{the application part described in the sub-sections 3.1-3.3
are worth to present in which we have also obtained present value
of the black body temperature of the universe}

In section 2, we have given thermodynamical approach to creation of
particles and have devloped field equations for thermodynamically
open homogeneous and isotropic spacially flat universe. In section
3, we developed radiation dominated model and then connects it to
the usual RW matter-radiation regime.Interestingly, using $\left[12,13\right]$,
We have also obtained present values of the black body temperature
of the universe.In the same section, we have also obtained the original
Prigogine's model.\textbf{ Although the model originates from instability
of vacuum rather than from a singularity,still there is a couple of
unavoidable singularities in the model.}

\section{Thermodynamical approach to creation of particles and field equations}

\subsection{Adiabatic usual Einstein's Equations.}

Let us consider the standard FRW metric

\begin{equation}
ds^{2}=dt^{2}-R^{2}\left(t\right)\left[\frac{dr^{2}}{1-kr^{2}}+r^{2}\left(d\theta^{2}+sin^{2}\theta d\phi^{2}\right)\right]\end{equation}

which represent spacially homogeneuos and isotropic universe. Let
our universe contains both matter and radiation in the form of perfect
fluid whose energy momentum tensor is as follows

\begin{equation}
T_{ij}=\left(p+\rho\right)u_{i}u_{j}-pg_{ij}\end{equation}

where

\begin{equation}
\rho=\rho_{m}+\rho_{r}\end{equation}

$\rho_{m}$ is matter density, $\rho_{r}$ is the radiation density
and

\begin{equation}
p=\frac{\rho_{r}}{3}\end{equation}

$p$ is pressure due to radiation.

The Einstein field equation 

\begin{equation}
G{}_{ij}=-\kappa T_{ij}\end{equation}

for FRW specially flat space time are, 

\begin{equation}
H^{2}=\frac{\kappa\rho}{3}\end{equation}

\begin{equation}
2\dot{H}+3H^{2}=-\kappa p\end{equation}

where H is the Hubble constant. These two equations combine to give
energy conservation law 

\begin{equation}
\dot{\rho}+3\left(\rho+p\right)H=0\end{equation}

the last equation leads to 

\begin{equation}
d\left(\rho V\right)+pdV=0\end{equation}

where V=$R^{3}$ is spacially three volume. The equation (9) describes
adiabatic thermodynamical energy conservation law for any arbitrary
comoving closed volume.

\subsection{Modified Einstein's equations for thermodynamical open universe.}

Contrary to the law$\left(9\right)$ if we regard the whole universe
as the open thermodynamical system in which number of the particles
in a given volume is no longer constant, then the conservation law
(9) will be read as follows

\begin{equation}
d\left(\rho V\right)+pdV-\frac{h}{n}d\left(nV\right)=0\end{equation}

where $n=$$\frac{N}{V}$ is the number density and $h=p+$$\rho$
is the enthalpy per unit volume.

Prigogine interpretated that the change in the number of the particles
in a volume is due to the tranfer of the energy from gravitation to
matter which means that creation of matter acts as source of the internal
energy. Moreover entropy change dS which vanishes for adiabatic transformation
in a closed system is given as follows for the adiabatic transformation
in a open system.

\begin{equation}
\frac{dS}{S}=\frac{dN}{N}\end{equation}

As the entropy of a system can not decrease we must have 

\begin{equation}
dN\geq0\end{equation}

This equation implies that energy in space time can produce matter
but the reverse process is thermodynamically forbidden.

Equation (10) can be written as 

\begin{equation}
\dot{\rho}-(\dot{n}/n)\left(\rho+p\right)=0\end{equation}

This corresponds to the equation $\left(8\right)$. Now as eq(8) is
obtained from (6) \& (7) so for open thermodynamical universe admitting
particle creation the Friedmann dynamical equation's need modification.Prigogine
modified equation(7) by adding a supplementary pressure $p_{c}$ to
its right hand side as follows

\begin{equation}
2\dot{H}+3H^{2}=-\kappa(p+p_{c})\end{equation}

where \begin{equation}
p_{c}=-\frac{h}{n}\frac{d\left(nV\right)}{dV}\end{equation}

$p_{c}$corresponds to creation of matter and it is negative or zero
depending on the presence or absence of particle creation.The modified
Einstein's equations are equations(6,13 \&14).

The equation$\left(13\right)$becomes

\begin{equation}
\left(\dot{\rho_{m}}+\dot{\rho_{r}}\right)=\left(\rho_{m}+\frac{4\rho_{r}}{3}\right)\frac{\dot{n}}{n}\end{equation}
 $ $ 

If matter and radiation co-exist without much interaction , this equation
splits into two parts 

\begin{equation}
\dot{\rho_{m}}=\rho_{m}\frac{\dot{n}}{n}\end{equation}

\begin{equation}
\dot{\rho_{r}}=\frac{4\rho_{r}\dot{n}}{3n}\end{equation}

which yields on integration

\begin{equation}
\rho_{m}=m_{b}n\end{equation}

\begin{equation}
\rho_{r}=\left(\rho_{r}\right)_{0}\left(\frac{n}{n_{0}}\right)^{\frac{4}{3}}\end{equation}

where $m_{b}$ is barion mass, $\left(\rho_{r}\right)_{0}$ is radiation
density and $n$$_{0}$ is number density at certain time $t=t_{0}$

\section*{3. Generalized Prigogine's models of the universe.}

As our model involve one extra function n (particle creation density)
beside usual$\rho$,p and R, we make following suitable assumption.

\begin{equation}
\frac{\dot{S}}{S}=\frac{\dot{N}}{N}=\alpha\geq0\end{equation}

wherw $\alpha$ is non negative constant. This means that enyropy
of universe varies exponentially with time. This is consistent with
equetion (12).The equ's(6,20 \&21) gives the following 

\begin{equation}
\left(\dot{R}/R\right)^{2}=H_{0}^{2}\left[\mu+\left(1-\mu\right)\left(n/n_{0}\right)^{1/3}\right]\left(n/n_{0}\right)\end{equation}

where

$\mu=m_{b}n_{0}/\rho_{0}$

and

$\rho_{0}=m_{b}n_{0}+(\rho_{r})_{0}$ is the total density at the
time $t=t_{0}.$

Now we discuss two special case.

\section*{3.1: Radiation dominated Model.}

The early universe was radiation dominated. There was no barionic
matter, so we take 

\[
\mu=0\;\;\&\;\;\rho_{r}\neq0\]

Equations (22) now becomes

\begin{equation}
\dot{R}/R=H_{0}\left[\left(n/n_{0}\right)^{2/3}\right]\end{equation}

Integrating equations (21\&23) we get the following solution

\begin{equation}
n=n_{0}\phi^{-3/2}\left(t\right)exp\left(\alpha t\right)\end{equation}

\begin{equation}
R=R_{0}\phi^{1/2}\left(t\right)\end{equation}

\[
H=H_{0}\phi^{-1}\left(t\right)exp\left(2\alpha/3\right)\]

\begin{equation}
\rho_{r}=\left(\rho_{r}\right)_{0}e^{4\alpha t/3}\phi^{-2}\left(t\right)\end{equation}

where

\begin{equation}
\phi(t)=\left[1+c\left(exp\left(2\alpha t/3\right)-1\right)\right],c=3H_{0}/\alpha\end{equation}

These solutions exhibit following properties 

(i) At \[
t=0,R=R_{0},H=H_{0},n=n_{0}\&\rho_{r}=\left(\rho_{r}\right)_{0}.\]

which are the cosmological quantities at the time $t=t_{0}$,so $t_{0}=0$
.This means that like prigogines model {[}1{]},the present model also
emerges without singularity with non zero particle creation density
$n_{0}$describing initial Minkowskian fluctuation. and others non
zero cosmological parameters.

(ii) We may take the initial values like $R_{0,}$etc as unity 1.
After a characteristic time \[
t_{c}=\frac{3}{2\alpha}\]

the model ultimately reaches to de-sitter regime given by

\begin{equation}
R_{d}\left(t\right)\simeq c^{1/2}e^{t/2t_{c}}=c^{1/2}e^{H_{d}t}\end{equation}

\begin{equation}
H_{d}\simeq\frac{\alpha}{3}=\frac{1}{2t_{c}}\end{equation}

\begin{equation}
n_{d}\simeq n_{0}\left(\alpha/3\right)^{3/2}\end{equation}

\begin{equation}
\left(\rho_{r}\right)_{d}\simeq\left(\rho_{r}\right)_{0}\alpha^{2}/9\end{equation}

\begin{equation}
\left(p_{r}\right)_{d}=(\rho_{r})_{d}/3\end{equation}

Where $\rho_{d}$ etc means density etc for radiation dominated de-sitter
regime.

\textbf{From the expression obtained for the scale factor, equation
(25) with (27), it is possible to say that there is no big bang, and
that the universe emerges from a vacuum state.However the function
$\phi$ in equation (27) can be zero for a fi{}nite time (negative
of course). We would say that the model contains a singularity}\textbf{\emph{. }}

\subsection*{3.2The usual RW radiation-matter dominated regime}

The de Sitter stage survives during decay time $t_{d}$ of its constituents
and then connects to a usual (adiabatic) matter-radiation RW universe
characterised by a matter-energy density $\rho_{b}$ and radiation
energy density $\rho_{\gamma}$ related to the RW funtion by 

\begin{equation}
\kappa\rho_{b}=3a/R^{3},\kappa\rho_{r}=3b/R^{4},\&\rho_{r}=\pi^{2}T^{4}/15\end{equation}

$a$ and $b$ are constants related to the total nomber $N_{b}$ of
baryons and photons $N_{r}$ in a volume $R^{3}$ and $T$ is the
black body radiation temperature. The connection at the decay time
$t_{d}$betveen desitter and matter radiation regimes fixes the constants
$a$ and $b$: we take

\[
a\simeq2H_{d}^{2}C_{r}^{2}exp\left(\left(5/3\right)H_{d}t_{d}\right)\]

\begin{equation}
b\simeq H_{d}^{2}C_{r}^{8/3}exp\left(4H_{d}t_{d}\right)\end{equation}

This implies that the (constant) specific entropy $S$ per proton
is 

\[
S=n_{r}/n_{b}\]
 where \[
n_{r}=2\zeta\left(3\right)T^{3}/\pi^{2},\&\:\; n_{b}=\rho_{b}/m_{b}\]

$\therefore$\begin{equation}
S=[\zeta\left(3\right)/3\pi^{2}][45/\pi^{2}]^{3/4}\kappa^{1/4}m_{b}\left(2t_{d}\right)^{1/2}e^{2t_{d}/3t_{c}}\end{equation}

The value of adiabatic invarient $\rho_{r}/T\rho_{b}$is

\begin{equation}
\rho_{r}/T\rho_{b}=b^{3/4}/a\left(\pi^{2}\kappa/45\right)^{1/4}\end{equation}

Hence these values are fixed by the knowledge of the two times $t_{c}$and
$t_{d}$. In the reference $\left[12\right]$ Prigogine and others
found quantum mechanically the values of $t_{c}$and $t_{d}$ in term
of mass $M$ produced. These are

\begin{equation}
t_{d}\simeq2.5\left(M/M_{p}\right)^{3}t_{p}\;\;\&\;\; t_{c}\simeq1.42\left(M/M_{p}\right)^{2}t_{p}\end{equation}

where $M_{p}$ and $t_{p}$ are the Plank mass and the Plank time
repectivily.In reference$\left[13\right]$Spindel. P. have found the
value of $M$ very close to the quantum mechanically produced one
$\left(53.3M_{p}\right)$ for correct observed of $S$ in between
$10$$^{8}to$$10^{10}.$ 

For \[
M/M_{p}=53.3\]
 we get from $\left(35\right)$ \[
S\cong10^{21}\]

\textbf{\textit{\emph{Here it should be noted the that the radiation
and matter dominated phases are re-obtained after a de Sitter regime.
With the help of the equation (34), we have imposed the transition
by hand. Even if the scale factor function is continous, it seems
the fi{}rst derivative is not continous. This can lead to a new singularities.}}}

\subsection*{3.3: The present black body radiation temperature.}

The present black body temperature is deduced from the continuity
requirement to be

\begin{equation}
T_{P}=\left(45/\pi^{2}\kappa\right)^{1/4}\left(b^{1/4}/a^{1/3}\right)H_{P}^{2/3}\end{equation}

\begin{equation}
T_{P}\left(^{0}K\right)\approx2.96\times10^{9}\left(H_{P}/75km/s/Mpc\right)^{2/3}\left(M/M_{P}\right)^{1/3}\end{equation}

where $H_{P}$ is the present observed value of the Hubble's constant
the observed black body radiation temperature$(2.70$$^{0}K)$ is
also obtained by taking $\left(M/M_{P}=50\right)$

From $\left(39\right)$

\begin{equation}
T_{P}=3.66(H_{P}/75km/s/Mpc)^{(2/3)}{}^{0}K\end{equation}

\subsection*{3.4: The Prigogine's dust filled model}

We take $\mu=1$ in equation(22) which means that $\rho_{r}=0$ and
the universe becomes dust filled. Equations(21-22) yields the following
solutions on integration.

\[
R^{1/2}\dot{R=H_{0}R_{0}^{2/3}e^{\alpha\left(t/2\right)}}\]

\[
R=R_{0}\phi^{2/3}\left(t\right)\]

\[
H=H_{0}e^{\alpha t/2}\phi^{-2}\left(t\right)\]

\[
n=n_{0}e^{\alpha t}\phi^{-2}\left(t\right)\]

\[
\rho_{m}=\left(\rho_{m}\right)_{0}e^{\alpha t}\phi^{-2}\left(t\right)\]

Where 

\[
\phi\left(t\right)=\left[1+c\left(e^{\alpha t/2}-1\right)\right]\]

and

\[
c=3H_{0}/\alpha\]

We note that if the constant $\alpha$ be replaced by $\alpha\kappa M/3$,
then we get a solution exactly similar to Prigogine's one$^{1}$.
This justifies that our model is a generalized Prigogine's models.

\end{document}